\documentclass{aa}
\usepackage{graphics}
\begin{document}

\thesaurus{(08.01.2; 08.03.5; 08.06.1; 08.09.2; 13.25.5)}

\title{The active binary star II\,Peg with {\it Beppo}SAX}

\author{S. Covino\inst{1} 
	\and G. Tagliaferri\inst{1}
          \and R. Pallavicini\inst{2}
 	\and R. Mewe\inst{3}
 	\and E. Poretti\inst{1}}
\offprints{S. Covino}
\institute{Brera Astronomical Observatory, Via Bianchi 46, I-23807 Merate, 
	Italy
	\and Palermo Astronomical Observatory, P.za del Parlamento 1, 
	I-90134 Palermo, Italy
          \and SRON Laboratory for Space Research, Sorbonnelaan 2,
          3584 CA Utrecht The Netherlands}

\date{Received date / accepted date}

\titlerunning{The active binary star II\,Peg}
\authorrunning{Covino et al.\ }

\maketitle

\begin{abstract}
\object{II\,Peg} is an ideal target to study stellar activity 
and flares, since intense and long lasting flares have been frequently
detected from this system at all wavelengths. We report here about a
{\it Beppo}SAX observation of \object{II\,Peg}. We followed the system
for $\sim 19$ hours on December 5 and 6 1997 with {\it Beppo}SAX and the
X-ray light curve resembles the typical behavior of a decay phase of a
long-lasting flare. The spectral analysis shows that the \object{II\,Peg}
X-ray spectrum is described by a two--temperature components, with the
two dominant temperatures centered in the range of 9--11 and 24--26\,MK.
The derived coronal metal abundance is low ($ Z \sim 0.2 Z_\odot$) 
compared to recent determinations of the photospheric abundance 
($Z \sim 0.6 Z_\odot$).
Some possible explanations for this phenomenology are reviewed. As for 
most other stellar coronal sources observed with {\it Beppo}SAX, we find that
in order to fit the {\it Beppo}SAX spectra an interstellar column density 
about a factor ten higher than previously determined is required.
\end{abstract}

\section{Introduction}
\label{sec:introduction}

The Italian-Dutch satellite {\it Beppo}SAX, thanks to the wide energy range 
covered by its detectors, has led to important contributions in the field of 
stellar corona phenomenologies. Indeed, hard X-ray emission above 10\,keV  has 
been, for the first time, detected during strong flares from \object{UX\,Ari} 
(Pallavicini \& Tagliaferri \cite{PT98}), \object{Algol} (Favata \cite{F98}) 
and \object{AB\,Dor} (Pallavicini et al. \cite{PTM99}). These observations 
have also allowed time-resolved spectroscopy of flares and the determination
of coronal metallicities on a wide energy band (from 0.1 to more than 10\,keV,
see, e.g. Favata \cite{F98} and Pallavicini et al. \cite{PTM99} for reviews of 
early {\it Beppo}SAX results).
 
With this aims in mind, we have observed with {\it Beppo}SAX the active 
RS CVn star \object{II\,Peg} (\object{HD224085}; \object{BD\,+27\,4642}), a 
single--line spectroscopic binary with a period of 6.7 days composed of a 
primary of K2 IV-V spectral type and an unobserved companion (Vogt 1981). The 
distance has normally been assumed to be $\sim 30 \ {\rm pc}$, as derived by 
the parallax of 0.034'' (Jenkins \cite{J63}, Vogt 1981), however the new 
{\it Hipparcos} satellite measurements give a value of $d = 42 \pm 1.6$\,pc 
($23.62 \pm 0.89$\,mas). 
This system (global $V \sim 7.62$; $B-V \sim 1.01$) shows luminosity 
variations up to 0.5\,mag presumably due to rotational modulation (see, for 
instance, Berdyugina et al. \cite{BBI98} and O'Neal et al. \cite{NSN98}). A 
detailed study of the stellar and orbital features of this system has been 
recently reported by Berdyugina et al. (\cite{BJITF98}).

Intense flares have been observed on \object{II\,Peg} from the the radio 
to the optical, UV and X-ray bands. This flaring activity is particularly
pronounced at higher frequencies;
flares have been observed by {\it Ariel V} (Schwartz \cite{SGR81}), in all 
IUE runs in 1981, 1983, 1985 and 1986 (Doyle et al. \cite{DBO89}) and also in 
the EUV band with the EUVE satellite (Mewe et al. \cite{MKOV97}). In the 
X-rays, three strong flares have been observed with EXOSAT, GINGA and ASCA, 
with an energy release of $>2 \times 
10^{35}\ \mbox{erg}$ in the first two cases. The flare seen by EXOSAT in the 
0.05-7\,keV energy band lasted longer than a day, with a rise time of 
$\sim 3$ hours, a peak phase of at least 2 hours and a long decay 
unfortunately not continuously observed by EXOSAT because of perigee 
passages. The quiescent value was reached again only two days after the 
occurrence of the peak of the flare (Tagliaferri et al. \cite{TWDCHS91}). 
The intense 
GINGA flare has been only partially observed during the decay phase, however
it was still so intense to be detected up to 18\,keV, with a power-law tail 
at energies greater than 10\,keV (Doyle et al. \cite{DOK92}). Finally, 
the ASCA flare lasted for about 18 hours, with an enhancement of a factor of 
$\sim 4$ in the iron abundance during the rise phase of the flare (Mewe 
et al. \cite{MKOV97}) and a total energy of $2.7 \times 10^{34}$\,erg. 

Quiescent X-ray emission from \object{II\,Peg} has also been observed by 
the ROSAT PSPC during the ROSAT All-Sky Survey (RASS, Dempsey et al. 
\cite{DLFS93a} and \cite{DLFS93b}) and as a pointed target (Huenemoerder \& 
Baluta \cite{HB98}).
The RASS observation showed a luminosity of $\sim 6 \times 
10^{30}$\,erg\,s$^{-1}$ with a bimodal temperature distribution with a high 
and a low temperature component of $2 (\pm 0.5)$ and $21 (\pm 5)$\,MK, 
respectively. The emission measure ratio, 
${\rm EM}_{\rm low}/{\rm EM}_{\rm high}$ was $\sim 
0.06$ and no interstellar absorption was constrained by the data (Dempsey 
et al. \cite{DLFS93b}). The pointed observation lasted 60\,Ksec for an eventual 
18\,Ksec effective observing time. The best-fit bimodal temperature 
distribution gave a luminosity of $\sim 7 \times 10^{30}$\,erg\,s$^{-1}$, 
high and low temperature components of $13 (\pm 1)$ and $\sim 3 (\pm 0.2)$\,MK 
with ${\rm EM}_{\rm low}/{\rm EM}_{\rm high} \sim 0.33$ and interstellar 
absorption of $N_{\rm H} = 7 (\pm 5) \times 10^{18}$\,cm$^{-2}$. 
Eventually, the best--fit coronal metal abundance turned out to be 0.3-0.4 
$(\pm 0.1)$\,Solar.

Since both the EXOSAT and GINGA flares were observed during the minima
of the photometric wave, and a large number of UV and optical flares have
been observed at the same phase, we have followed \object{II\,Peg} for $\sim 
19$ hours close to the minimum phase, on 1997, December 5 and 6, with the 
purpose of detecting an intense flare as those observed by EXOSAT, 
GINGA and ASCA. Moreover, since simultaneous multi-wavelength observations of 
stellar flares are extremely important to study the physical processes that 
drive thermal and non-thermal coronal plasmas (e.g. Haisch \& Rodon\`o 
\cite{HR89} and references therein), we arranged for simultaneous optical 
observations in the standard B and V filters.

This paper is organized as follows. In Sect.\,\ref{sec:data} we describe 
the data sample and the analysis technique applied. In Sect.\,\ref{sec:results}
the results are presented, and they are finally discussed in 
Sect.\,\ref{sec:discussion}.

\section{The Data}
\label{sec:data}

The {\it Beppo}SAX satellite carries aboard a number of X-ray detectors 
able to cover a wide energy range, from 0.1 to 300\,keV (Boella et al. 
\cite{BBPPSB97}). Actually, apart from some very intense X-ray flares,
the study of coronal sources can only be performed with the Low Energy 
Concentrator Spectrometer (LECS, Parmar et al. \cite{PMB97}) and the two 
Medium Energy Concentrator Spectrometers (MECS, Boella et al. \cite{BCC97}). 
Still these instruments are sensitive to a large energy range. In particular
the LECS covers the range from 0.1 to 10 keV, with a spectral resolution 
comparable at the lowest energies to that of CCD detectors (such as those 
used on ASCA). Favata et al. (\cite{FMPS97}) have argued on the basis of
spectral simulations of ASCA and {\it Beppo}SAX LECS spectra that the broader
spectral band of the latter instrument should allow a better determination
of the overall coronal metallicity, even if the lower resolution at higher
energies makes it difficult to determine the individual metal abundances.
The MECS detectors cover only the 1.7--10\,keV energy range but they
have an effective area about two times larger than that of the LECS, 
allowing the study of the Fe\,K complex at $\sim 6.7$\,keV more effectively.

{\it Beppo}SAX observed \object{II\,Peg} from December 5 to 6 1997 for about
19 hours (observation sequential number 10161001), yielding $\sim 15$\,ks and
$\sim 35$\,ks of effective observing time in the LECS and MECS detectors, 
respectively. The difference in exposure time is due to the fact
that the LECS
can only operate when in the Earth shadow. 
The data analysis has been based on the linearized, cleaned, event files, 
obtained by the {\it Beppo}SAX Science Data Center (SDC; Giommi \& Fiore 
\cite{GF97}) on-line archive.
 
The light curves and spectra were extracted using the FTOOLS (v. 4.0)
package, with an extraction region of 8\arcmin.5 and 4\,\arcmin\ radius
for the LECS and the MECS, respectively. The different
regions are due to a broader Point Spread Function (PSF) for the LECS
at low energies, while above 2\,keV the LECS and MECS PSFs are
essentially comparable. The considered extraction regions contain more than 
90\% of the counts for all energies. 
The light curve analysis has been performed with the XRONOS (v.\,4.02) 
package,
while for the spectral analysis we used the XSPEC (v.\,10.0) package, 
with the response matrices released by the SDC in September 1997. 
The spectra
were rebinned following the recipe provided by the SDC. The rebinning 
samples the instrumental resolution with the same number of 
channels, three in our case, at all energies.
LECS and MECS background spectra were
accumulated from blank fields available at the public SDC ftp site (see Fiore
et al. \cite{FGG99}; Parmar et al. \cite{POO99}). 
LECS data have been analyzed only in the 0.1--4\,keV range due to still 
unsolved calibration problems at higher energies (Fiore et al. \cite{FGG99}). 
The LECS and MECS spectra were always conjunctly fit after having allowed a 
rescaling factor for LECS data in order to take into account the 
uncertainties in the inter-calibration of the detectors. The rescaling factor
turned out to be $\sim 0.75$, which is within the acceptable
range of $0.7 \div 1$ (Fiore et al. \cite{FGG99}). Therefore, we kept
this constant value fixed to 0.75 in all our spectral analyses. For a full 
description of the analysis techniques of {\it Beppo}SAX data see Fiore et 
al. (\cite{FGG99}).

As mentioned above, interesting spectral information could also be 
obtained in some cases from the Phoswich Detector System (Frontera et al. 
\cite{FCF97}) aboard {\it Beppo}SAX, even if this detector had been designed 
to provide the best sensitivity at energies relatively higher than those typical
for a stellar corona. The energy range covered by the PDS is from 15 to 
300\,keV. Hard X-ray emissions ($> 20$\,keV) have been detected during 
strong flares from active stars (Pallavicini et al. \cite{PTM99}). During
the present \object{II\,Peg} observation, however, no convincing evidence of a 
PDS detection has been found. This is not surprising since coronal sources
have been detected with the PDS only during the rise and the peak phase of
strong X-ray flares, and not during the decay phase (which was the only
one observed for \object{II\,Peg}).
   
The optical observations were performed in the standard BV filters,
using the Marcon 50-cm telescope of the Brera Astronomical Observatory
at Merate (Italy), equip\-ped with a photon-counting photometer (EMI\,9789QA).
\object{HR\,9088} (\object{85\,Peg}) and \object{HR\,8997} (\object{78\,Peg})
were used as comparison stars.

\section{Results}
\label{sec:results}

\begin{figure}
\begin{center}
\begin{tabular}{c}
{\resizebox{\hsize}{!}{\includegraphics{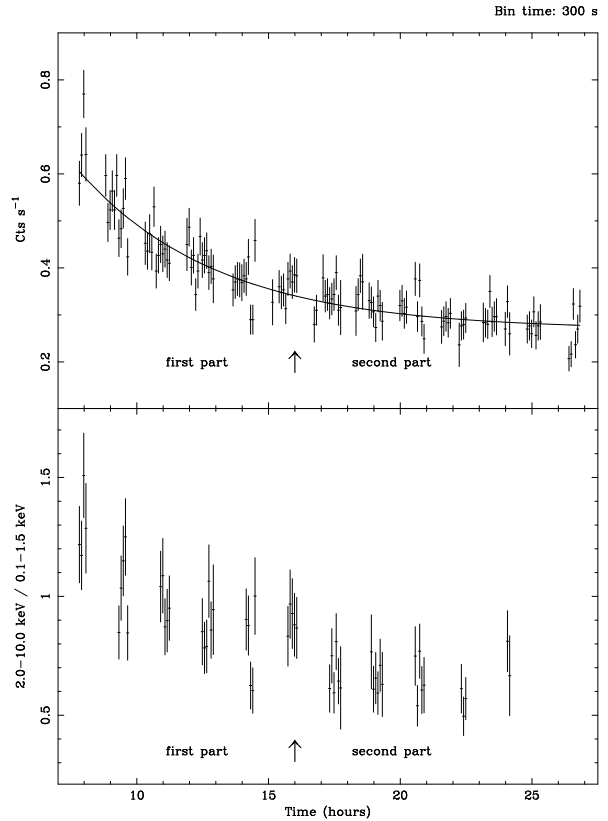}}}
\end{tabular}
\end{center}
\caption{Top panel. \object{II\,Peg} MECS light curve in the energy range 
from 2 to 10\,keV. The solid line shows the best fit exponential plus constant 
curve. The quiescent flux at the end of the observation turns out to be $\sim 
0.27$\,Cts\,s$^{-1}$ and the decay time $\sim 5.3$\,hours. The background has 
been stable during the observation and amounts to roughly one hundredth of the 
source flux in the MECS cameras. Bottom panel. Hardness ratio between 
2.0-10.0\,keV (MECS) data and 0.1-1.5\,keV (LECS) data. The binning time is 
300\,s. The observation started on December 5 1997 and the arrow signs the 
separation between the first and the second part of the observation considered 
in the spectral analysis.}
\label{fig:lc}
\end{figure}

In Fig.\,\ref{fig:lc} we show the total \object{II\,Peg} MECS light curve
and the 2.0-10.0\,keV / 0.1-1.5\,keV (MECS/LECS) hardness ratio. 
The best fit exponential plus constant fit of the MECS light curve is also 
shown. The quiescent value determined at the end of the observation turns
out to be $\sim 0.27$\,Cts\,s$^{-1}$ while the flare decay time is $\sim
5.3$\,hours. The background amounts only to roughly one hundredth 
of the source flux and has been stable for the whole observation. 
Variability is clearly present, the flux in the $2 \div 10$\,keV band 
decreases by a factor $\sim 2$ in about 10 hours. This behavior 
would suggest that we are observing the tail of a long-lasting flare. 
Unfortunately, the optical observations do not help in this respect since 
during the {\it Beppo}SAX observation bad weather at Merate prevented us to 
perform any \object{II\,Peg} photometry. Observations 
could only be performed on the next night (December 6-7) when the source did 
not show any sign of activity. No variability larger than the observational 
uncertainties was detected; the standard deviations of the differential 
measurements \object{HD\,224016}-\object{II\,Peg} ($\pm 0.01$\,mag in both 
colors) are similar to those measured between the comparison stars.

For the spectral analysis, we have applied the thin plasma models by 
Raymond \& Smith (\cite{RS77}; RS) and Mewe et al. (\cite{MKL96}; MEKAL) as
implemented in XSPEC. No significant differences have been found in 
the results obtained with the two models and therefore we report here only 
the MEKAL results.
The errors on the counts have been taken into account
by the Gehrels (\cite{G86}) approximation for data following the Poissonian
statistics, rather than the less adequate Gaussian statistics. Our data 
are always in the Gaussian regime but for energies above $\sim 6$\,keV for the
MECS cameras, below $\sim 0.2$\,keV and around $\sim 0.4$\,keV for the LECS
camera. The $\chi^2$ minimization statistics was applied throughout the paper.

Abundance variations in the source spectrum have been modeled through the use 
of a single parameter, the global ``metallicity'' $Z$, by assuming a fixed
ratio between individual elemental abundances and the corresponding solar 
photospheric values as given by Anders \& Grevesse (\cite{AG89}).

The interstellar absorption $N_{\rm H}$ has been included in the fit. At first,
we have allowed $N_{\rm H}$ to vary freely in the fit procedure and the 
best--fit value was $\sim 8 \times 10^{19}$ cm$^{-2}$ 
(see Table\,\ref{tab:spec2}). 
This value is more than a factor of ten higher then the one obtained by the 
analysis of ROSAT (Huenemoerder \& Baluta \cite{HB98}), EXOSAT and EUVE 
data (Ta\-glia\-ferri et al. \cite{TWDCHS91}; Mewe et
al. \cite{MKOV97}) where the derived interstellar reddening amounted
to $5-8 \times 10^{18}$ cm$^{-2}$. Indeed, applying the relation $H \sim 
0.07\,\mbox{cm}^{-3}$ (Paresce \cite{P84}) a value of 
$N_{\rm H} \sim 9 \times 10^{18}\,\mbox{cm}^{-2}$
is obtained, rather close to the value obtained by the analyses
of ROSAT, EXOSAT and EUVE data and still an order of magnitude lower 
than the result obtained by the fit of {\it Beppo}SAX data.

\begin{table*}
\begin{center}
\begin{tabular}{|c|cccccccc|}
\hline
& $N_{\rm H}$ & $KT_1$ & $KT_2$ & Z/Z$_\odot$ & $\frac{\mbox{EM}_1}{\mbox{EM}_2}$ & flux$_{\mbox{0.1-2.4\,KeV}}$ & d.o.f. & $\chi^2$ \\
& $10^{19}$\,cm$^{-2}$ & (keV)  & (keV)  &	      &                     & $\frac{\mbox{erg}}{\mbox{s cm}^2}$       $\times10^{-11}$      &        &          \\
\hline
{\bf Tot. Sample} 
& $7.8\pm_{2.8}^{2.2}$ & $1.04\pm_{0.09}^{0.09}$ & $2.55\pm_{0.11}^{0.11}$ & $0.16\pm_{0.04}^{0.04}$ & 0.60 & 5.1 & 86 & 1.0 \\ 
\hline  
{\bf 1st Part}   
& $8.3\pm_{3.3}^{4.2}$ & $1.05\pm_{0.14}^{0.16}$ & $2.68\pm_{0.15}^{0.38}$ & $0.17\pm_{0.06}^{0.06}$ & 0.44 & 5.3 & 86 & 0.8 \\ 
\hline  
{\bf 2nd Part}  
& $7.8\pm_{5.7}^{5.4}$ & $1.03\pm_{0.48}^{0.12}$ & $2.45\pm_{0.59}^{1.06}$ & $0.14\pm_{0.05}^{0.14}$ & 1.02 & 4.9 & 86 & 1.1 \\ 

\hline
\end{tabular}
\end{center}
\caption{Best fit parameters with two--temperature MEKAL models.  
Spectra have been analyzed considering both the whole data set and the first 
and second part of the observation alone (see also Fig.\,\ref{fig:lc}). The 
interstellar reddening $N_{\rm H}$ was free to vary during the fits, the fluxes 
are corrected for the absorption. Assuming a 42\,pc distance, the 
luminosity turns out to be $\sim 10^{31}$\,erg\,s$^{-1}$. The 
errors on the parameters are at 90\% for four parameters of interest. The 
fits are satisfactory but $N_{\rm H}$ is an order of magnitude larger than 
expected.}
\label{tab:spec2}
\end{table*}

In order to minimize the number of free parameters and also the effect that
$N_{\rm H}$ could have on the derived metallicity (a higher $N_{\rm H}$ could 
be compensated by a low metal abundance in the fit procedure),
we have also fitted our data frozing
$N_{\rm H}$ to $5 \times 10^{18}$ cm$^{-2}$.
The results are reported in Table\,\ref{tab:spec1}.

\begin{table*}
\begin{center}
\begin{tabular}{|c|ccccccc|}
\hline
& $KT_1$ & $KT_2$ & Z/Z$_\odot$ & $\frac{\mbox{EM}_1}{\mbox{EM}_2}$ & flux$_{\mbox{0.1-2.4\,KeV}}$ & d.o.f. & $\chi^2$ \\
& (keV)  & (keV)  &	      &                     & $\frac{\mbox{erg}}{\mbox{s cm}^2}$       $\times10^{-11}$      &        &          \\
\hline
{\bf Tot. Sample} 
& $0.90\pm_{0.12}^{0.18}$ & $2.32\pm_{0.11}^{0.17}$ & $0.32\pm_{0.06}^{0.08}$ & 0.19 & 4.6 &87 & 1.8 \\ 
\hline  
{\bf 1st Part}
& $0.90\pm_{0.19}^{0.34}$ & $2.47\pm_{0.15}^{0.35}$ & $0.33\pm_{0.09}^{0.11}$ & 0.13 & 4.8 & 87 & 1.4 \\ 
\hline  
{\bf 2nd Part}
& $0.87\pm_{0.27}^{0.33}$ & $2.11\pm_{0.20}^{0.41}$ & $0.31\pm_{0.10}^{0.14}$ & 0.16 & 4.4 & 87 & 1.3 \\ 
\hline 
\end{tabular}
\end{center}
\caption{Best fit parameters with two--temperature MEKAL models. Spectra
have been analyzed considering both the whole data set and the first and 
second part of the observation alone (see also Fig.\,\ref{fig:lc}). The 
$N_{\rm H}$ value has been frozen to $5 \times 10^{18}$ cm$^{-2}$, the fluxes 
are corrected for the absorption. Assuming a 42\,pc distance, the 
luminosity turns out to be $\sim 9 \times 10^{30}$\,erg\,s$^{-1}$.
The errors on the parameters are at 90\% for three parameters of
interest. The fits are worse than those in Table\,\ref{tab:spec2} 
and formally unacceptable.}
\label{tab:spec1}
\end{table*}

\begin{figure}
\begin{center}
\begin{tabular}{c}
{\rotatebox{270}{\resizebox{6.5truecm}{!}{\includegraphics{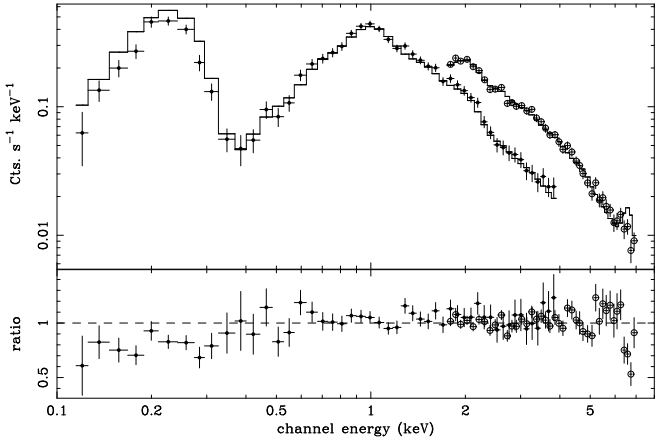}}}}
\end{tabular}
\end{center}
\caption{LECS+MECS spectra for the whole data set with $N_{\rm H}$ frozen to
$5 \times 10^{18}$ cm$^{-2}$. The ratio between the data and the MEKAL 
two--temperature best--fit model are shown in the lower panel.}
\label{fig:totnhfix}
\end{figure}

Satisfactory fits to the data were not possible with single--temperature 
models. Two--temperature models with $N_{\rm H}$ fixed to 
$5 \times 10^{18}$ cm$^{-2}$ were also not able to provide satisfactory fits 
to the data and strong residuals are clearly visible below 0.4\,keV 
(Fig.\,\ref{fig:totnhfix}). 
To check if this result could be related to the source variability the
observation was subdivided in two distinct parts (Fig.\,\ref{fig:lc}), 
and LECS and MECS spectra were accumulated for the first and second part
separately. However, even in these two cases we did not obtain any 
satisfactory fits with this $N_{\rm H}$ value (Table\,\ref{tab:spec1})
and the results for the two time intervals were essentially indistinguishable.

\begin{figure}
\begin{center}
\begin{tabular}{c}
{\rotatebox{270}{\resizebox{6.5truecm}{!}{\includegraphics{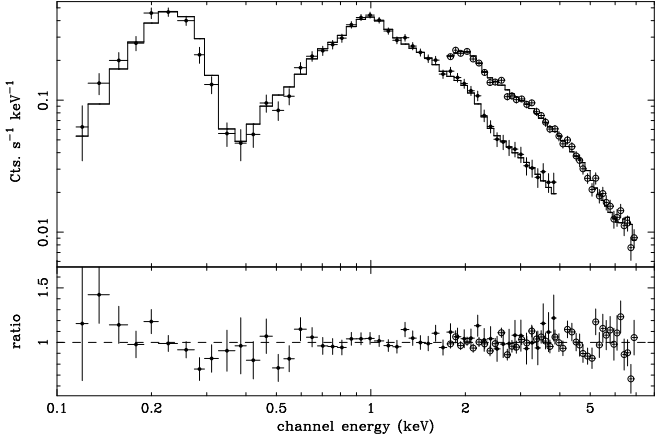}}}}
\end{tabular}
\end{center}
\caption{LECS+MECS spectra for the whole data set with $N_{\rm H}$ free to 
vary. The ratio between the data and the MEKAL two--temperature best--fit model
are shown in the lower panel.}
\label{fig:totnhfree}
\end{figure}

In all three cases significant improvement to the fits are obtained by 
two--temperature models with $N_{\rm H}$ and the metal abundance free to vary.
In these cases the metal abundance value is lower and more similar to the
value found from the ROSAT (Huenemoerder \& Baluta \cite{HB98}), ASCA and EUVE 
data (Mewe et al. 1997). The best--fit 
models are shown in Figures\,\ref{fig:totnhfix} and \ref{fig:totnhfree} 
and refer to two--temperature MEKAL models for the whole data set with 
$N_{\rm H}$ either frozen to $N_{\rm H} = 5 \times 10^{18}$ cm$^{-2}$ or 
free to vary. Consequently, our {\it Beppo}SAX data can be only fitted 
assuming a large $N_{\rm H}$. A similar result was obtained for 
{\it Beppo}SAX observations of the coronal sources \object{VY\,Ari}, 
\object{HD\,9770}, \object{UX\,Ari}, \object{AB\,Dor} and \object{AR\,Lac} 
(Favata et al. 1997b, Tagliaferri et al. \cite{TCCP99}, Palla\-vicini \& 
Tagliaferri \cite{PT98}; Palla\-vicini et al. \cite{PTM99}, Rodon\`o et al. 1999).
This anomaly may be due either to a problem in the calibration of the
LECS detector below 0.5\,keV (i.e. where $N_{\rm H}$ is estimated in case of 
values lower than $\sim \times 10^{20}$ cm$^{-2}$) or to an incorrect source 
modeling evidenced by the wide {\it Beppo}SAX energy range.

Two-temperature fits performed removing the energy channels below 0.4\,keV 
show that any interstellar absorption (up to few $\times 10^{20}$\,cm$^{-2}$) 
is adequate to fit the data (i.e. $N_{\rm H}$ is no more constrained). The fits 
are always statistically acceptable ($\chi^2 \sim 1.1 \div 1.2$) and the 
metal abundance remains unchanged ($Z \sim 0.2\,Z_\odot$).
On the contrary, the data can not be fitted with solar abundance even with 
the addition of a third temperature component. This is also true if we use the 
\object{II\,Peg} photosperic abundance ($Z = 0.6\,Z_\odot$) derived by Ottmann
et al. (1998).

\section{Discussion}
\label{sec:discussion}

The EXOSAT, GINGA, ROSAT, EUVE and ASCA observations of \object{II\,Peg},
have shown that its quiescent
coronal emission is characterized by a hot plasma, with temperatures up to
20 million degrees. In particular Mewe et al. (1997) using 
EUVE and ASCA data determined the differential emission measure (DEM)
distribution of this source. They found a bimodal distribution for the data
of both satellites, with the two peaks centered at $\sim 10$ and
$\sim 20$ MK for the ASCA data. In both cases the DEM analysis
gave results fully consistent with the one obtained using a two--temperature 
fit. Our {\it Beppo}SAX results are similar to the
ASCA ones, we essentially find the same two temperatures (see 
Table\,\ref{tab:spec2}), but the ASCA softer EM is larger than the harder 
one, while our harder EM is a factor of two larger than the softer one.
This could be due to the decay phase of the flare, that we seem to have 
detected. However, our results are also 
essentially in agreement with those derived by the ROSAT data analysis of 
Huenemoerder \& Baluta (\cite{HB98}) once we take into account the small ROSAT 
energy band-width. In that case the ROSAT light curve only shows a steady 
linear decline from the beginning of the observation of $\sim 20\%$ without
significant evidence of flare activity.
The {\it Beppo}SAX data also show
a low coronal metal abundance, significantly lower than solar, 
again consistent with the ROSAT (Huenemoerder \& Baluta (\cite{HB98}), ASCA and 
EUVE results (Mewe et al. \cite{MKOV97}).

ASCA observations of active coronal stars systematically yield subsolar
metal abundances (White et al. \cite{WAD94}, White \cite{W96}; 
Singh et al. \cite{SDW95}, \cite{SWD96}; Tagliaferri et al. \cite{TCF97}, 
Ortolani et al. \cite{OMP97}, Mewe et al. \cite{MKWP96}, \cite{MKOV97}), 
and subsolar metal abundances have also been found by other X-ray 
satellites (Tsuru et al. \cite{TMO89}; Stern et al. 
\cite{SUTN92}, \cite{SLSP95}; Ottmann \& Schmitt \cite{OS96}; Schmitt et al. 
\cite{SSDK96}; Mewe et al. \cite{MKWP96}, \cite{MKOV97}). Similar results
are now obtained with {\it Beppo}SAX (Favata et al. \cite{FMPC97},
Tagliaferri et al. \cite{TCCP99}; Pallavicini \& Tagliaferri \cite{PT98};
Pallavicini et al. \cite{PTM99}; Rodo\-n\`o et al. 1999). 
This does not imply necessarily that these values are in contradiction 
with the photospheric abundances for the same stars. 
Indeed, for some of them they are not, but for some other stars
the derived coronal metallicities are significantly lower than the
photospheric values (see discussion in Tagliaferri et al. 1999 for 
a list of cases). The latter seems to be the case for \object{II\,Peg},
for which Ottman et al. (1998) from a detailed and sophisticated
analysis found recently a photospheric value of 
[Fe/H]\footnote{[Fe/H]=log [(Fe/H)/(Fe/H)$_\odot$]}
$\sim -0.2$, while we found a coronal metallicity that is a factor of 
three lower (see Table\,\ref{tab:spec2}). 

In their study, Ottmann et al. (\cite{OPG98}) have determined the 
photospheric metal abundances of several active stars in order to investigate 
on the possible abundance stratification in stellar atmospheres.
Their results for various RS CVn binaries (\object{AY\,Cet}, \object{VY\,Ari}, 
\object{EI\,Eri}, \object{IM\,Peg}, \object{$\lambda$\,And} and 
\object{II\,Peg}) show that all these stars have abundances $[Fe/H] > -0.4$.
In particular, for \object{II\,Peg} they found [Fe/H]$ \sim -0.2$ (i.e. $Z 
\sim 0.6\,Z_\odot$), in contrast with a previous determination by Randich et al. 
(\cite{RGP94}) who gave [Fe/H]$ = -0.5$ (i.e. $Z \sim 0.3\,Z_\odot$). If the value found 
by Ottmann et al. (\cite{OPG98}) is correct, we have a discrepancy by a factor 
$\sim 3$ between the photospheric and the coronal metallicities, similarly to
what has been reported for other active stars (e.g. \object{AB\,Dor}, Mewe et
al. \cite{MKWP96}). 
To explain the observed metallicity gradient two possible mechanisms are 
often invoked: the so called First Ionization Potential (FIP) effect and the 
hydrostatic equilibrium stratification. 

The FIP-effect has been invoked to explain the observed abundance gradient 
in the Solar atmosphere. The elements with a high FIP (N, O, Ne, Ar) are
underabundant in the corona, with respect H, by a factor of 3-4 compared to
the photosphere. On the contrary, elements with a low FIP (Fe, Mg, Si, Ca,
Na) are overabundant (Meyer \cite{M85}, Feldman \& Widing \cite{FW90}).
In case the FIP-effect is effectively working in \object{II\,Peg}, we should 
expect that overall metallicity $Z$ (which is essentially determined
by Fe) is enhanced (and not depleted, as observed) with respect to the 
photospheric value. Moreover, elements such as Fe, Mg and Si, which have the 
same FIP ($7 \div 8$\,eV), should have approximately the same ratio between 
their coronal and photospheric abundances.
Our {\it Beppo}SAX data do have not enough resolution to constrain the 
abundances of individual elements, but the coronal abundances derived from
the ASCA and EUVE observations by Mewe et al. (\cite{MKOV97})
do not appear to confirm this expectation in the case of \object{II\,Peg}.
In fact, whereas Fe, Mg and Si have similar photospheric abundances 
(Ottmann et al. \cite{OPG98}), the Fe and Si coronal abundances reported
by Mewe et al. (\cite{MKOV97}) differ by a factor of 2. If the coronal 
abundances derived by us and others from X-ray observations are correct, and
do not depend significantly on uncertainties on the atomic data and on the 
modeling (including neglected possible line opacity effects, cf. Schrijver
et al. \cite{SMVK95}), and if the photospheric abundances derived by Ottmann et 
al. (\cite{OPG98}) are also correct, there is a clear discrepancy in 
\object{II\,Peg} between photospheric and coronal abundances. However, the
FIP-effect is unlikely to be the cause for this discrepancy.

The second scenario, proposed by Mewe et al. (\cite{MKOV97}, see also van den 
Oord \& Mewe, \cite{OM99}), and widely discussed by Ottmann et al. 
(\cite{OPG98}), consists of the possibility that a reduction in the metal 
coronal abundance is simply the effect of the establishment of hydrostatic 
equilibrium. The scale height for each ion depends on the mass and charge and 
the depletion of the coronal abundances compared to the the photospheric ones 
may even reach a factor within $2 \div 10$. The whole problem seems rather 
complex and is still lacking of a full and satisfactory theoretical treatment.
However, Ottmann et al. (\cite{OPG98}), neglecting the effect of the ion
charge, conclude that for \object{II\,Peg} an agreement between observations
and predictions for Fe, Mg and Si can be found in the sense that the 
correct direction of the effect is predicted.

An interesting alternative and/or complementary explanation 
for these observations has been proposed recently by Drake (\cite{D98}). 
The mechanism consists in the possibility that the depletion of the coronal
abundances is only apparent and actually due to the enhancement of the 
coronal He abundance. The gross result would be that of a lowering of the 
ratio between lines and continuum, just what is actually measured to derive 
the metal abundance from X-ray observations of stellar coronae.
In stellar coronae with temperature $\sim 10^7$\,K, in fact, the continuum
emission is driven by inelastic collisions between electrons and H and He 
nuclei. He in coronae can be more abundant than in the photosphere due to 
a fractioning of the composition analogous at the FIP-effect (see, for
instance, Meyer \cite{M93}). There is, moreover, the possibility that He
nuclei are braked due to an inefficient Coulombian drag in the wind 
acceleration phase (see Geiss \cite{G82}). 
To mimic the metal abundances we derived in this paper ($Z \sim
0.2 Z_\odot$, or [Fe/H]$ \sim -0.7$) would be necessary 
to assume that the ratio between the He and the H coronal abundance for
\object{II\,Peg} is more than a factor of ten higher than that in the Sun
(see Fig.\,1 in Drake \cite{D98}, i.e. an He abundance
comparable to or higher than that of the H). We tried to fit the X-ray spectra
assuming a coronal metal abundance equal to the photospheric one
determined by Ottmann et al. (\cite{OPG98}) and letting the He abundance free
to vary. However, we could not find a good fit to the data in this way.
Thus, this does not seem to be a valid explanation for the \object{II Peg} 
case.

The energy released by the flare we detected is not easy to compute. We 
can only derive a lower limit assuming that the beginning of the flare is 
coincident with the beginning of our observation. We also assume that the
best-fit parameters reported in Table\,\ref{tab:spec2} for the first part
of the observation are adequate for an average description of the spectral
features of the flare. In such a case, with a minimum flare duration of
$\sim 10$\,hours the total fluence released amounts to $\sim 1.9 \times 
10^{-6}$\,erg\,cm$^{-2}$. With a distance of $\sim 42$\,pc this translates 
in a total energy release of $\sim 4 \times 10^{35}$\,erg, to be compared with
the energy released by the quiescent flux of the star in the same amount of time
($\sim 3.7 \times 10^{35}$\,erg). 
The quiescent luminosity observed by {\t Beppo}SAX is of the order of 
$10^{31}$\,erg\,s$^{-1}$ (see Tables\,\ref{tab:spec2} and \ref{tab:spec1}). 
Indeed, the recent revision of the \object{II\,Peg} system distance by 
{\it Hipparcos} moved the system $\sim 50\%$ farther than previously 
assumed. This implies that previous estimates of luminosities and total energy 
release for \object{II\,Peg} were underestimate by a factor $\sim 2$. 
Taking this into account, the {\it Beppo}SAX observed luminosity is in agreement
with those recorded in the previous ROSAT PSPC observation (Dempsey et al. 
\cite{DLFS93a}, \cite{DLFS93b}; Huenemoerder \& Baluta \cite{HB98}). 
The statistics of our observation eventually do not allow us to perform any 
time-resolved analysis and the only partial coverage of the phenomenum prevent 
us to try to model the flare in order to derive more flare parameters and to 
discriminate between different theoretical scenarios.

A new {\it Beppo}SAX observation of \object{II\,Peg} lasting 150\,ks has 
been approved, with the aim of detecting a strong flare. However, even if a 
large flare is catched and modeling of the flare will become possible, this new 
observation will not solve the problem of the low coronal metallicities found 
for this source by different satellites and of their apparent discrepancy with 
the photospheric values. This is a topic that will better be addressed with 
future more sensitive,  and with better spectral resolution, X-ray missions 
like CHANDRA, XMM and ASTRO--E. With these missions, it will also be possible 
to verify if the high $N_{\rm H}$ values indicated by {\it Beppo}SAX are real 
or not, and to understand their still unexplained origin.

\begin{acknowledgements}
This research was financially supported by the Italian Space Agency. We thank 
the {\it Beppo}SAX Science Data Center for their support in the data analysis.
We also thank the anonymous referee for her/his valuable comments and 
suggestions that helped us to improve the final version of this paper.
\end{acknowledgements}

\end{document}